\documentclass[twocolumn,twocolappendix]{aastex6}

\usepackage{txfonts}

\usepackage{color,bm}

\newcommand{\eqref}[1]{(\ref{#1})}

\shorttitle{A condensation--coalescence cloud model for exoplanets}
\shortauthors{Ohno \& Okuzumi}

\begin{document}

\title{A Condensation--Coalescence Cloud Model for Exoplanetary Atmospheres: Formulation and Test Applications to Terrestrial and Jovian Clouds}

\author{Kazumasa Ohno and Satoshi Okuzumi}
\affil{Department of Earth and Planetary Sciences, Tokyo Institute of Technology, Meguro, Tokyo, 152-8551, Japan}

\begin{abstract}
A number of transiting exoplanets have featureless transmission spectra that might suggest the presence of clouds at high altitudes. A realistic cloud model is necessary to understand the atmospheric conditions under which such high-altitude clouds can form. In this study, we present a new cloud model that takes into account the microphysics of both condensation and coalescence. Our model provides the vertical profiles of the size and density of cloud and rain particles in an updraft for a given set of physical parameters, including the updraft velocity and the number density of cloud condensation nuclei (CCN). We test our model by comparing with observations of trade-wind cumuli on the Earth and ammonia ice clouds in Jupiter. For trade-wind cumuli, the model including both condensation and coalescence gives predictions that are consistent with observations, while the model including only condensation overestimates the mass density of cloud droplets by up to an order of magnitude. For Jovian ammonia clouds, the condensation--coalescence model simultaneously reproduces the effective particle radius, cloud optical thickness, and cloud geometric thickness inferred from Voyager observations if the updraft velocity and CCN number density are taken to be consistent with the results of moist convection simulations and Galileo probe measurements, respectively. 
These results suggest that the coalescence of condensate particles 
is important not only in terrestrial water clouds but also in Jovian ice clouds.
Our model will be useful to understand how 
the dynamics, compositions, and nucleation processes in exoplanetary atmospheres
affects the vertical extent and optical thickness of exoplanetary clouds via cloud microphysics.
\end{abstract}


\section{Introduction} \label{sec:intro}
Recent observations of the transmission spectra of exoplanets revealed that some 
hot Jupiters \citep[e.g.,][]{Pont13,Sing15,Sing16}, hot Neptunes \citep[e.g.,][]{Crossfield+13,Enrenreich14,Knutson14a,Dragomir15,Stevenson+16}, 
and super-Earth \citep[e.g.,][]{Bean10,Kreidberg14,Knutson14b} have featureless spectra. 
One interpretation of the featureless spectra is that these exoplanets 
have dust clouds that block starlight at high altitudes \citep[e.g.,][]{Seager00,Fortney05}.
Dust clouds are also believed to have the crucial impacts on the observed spectra of brown dwarfs whose effective temperature fall into the same range of exoplanets \citep[e.g.,][]{Saumon&Marley08}.
For example, the observations of brown dwarfs show
a blueward shift of spectral energy distributions during L/T transition 
that might suggest the sinking of condensate particles \citep{AM01,Burgasser+02,Marley+02,Saumon&Marley08}. Observations also suggest spectral variability that might imply the effect of cloud spatial distributions \citep{Buenzli+12,Yang+15,Yang+16}.

A realistic cloud model that predicts the size and spatial distributions of condensation 
particles for arbitrary atmospheric conditions is necessary to 
understand the atmospheric properties of both exoplanets and brown dwarfs.

The microphysics that governs the formation of clouds is highly complex, 
and there are at least two processes by which cloud particles can grow. 
The first process is the condensation of vapor onto particles in an adiabatically cooling updraft. 
In terrestrial water clouds and Jovian ice clouds, this process is responsible for the growth of small 
particles to 10 $\micron$ in radius \citep[e.g.,][]{Rossow78}. 
Further growth of the particles proceeds through the second process, 
the coalescence driven by the differential settling under gravity.
This second process is essential for the initiation of precipitation in terrestrial water clouds \citep{Pru97}.
  
However, previous models of clouds in exoplanets as well as in brown dwarfs neglected or at least 
parametrized coalescence.
The convective cloud model by \citet{AM01}, which has been used by \citet{Morley13,Morley15}
for modeling the transmission spectrum of super-Earth GJ 1214b,  
encapsulates the effects of particle growth due to condensation and coalescence in 
a single free parameter $f_{\rm sed}$.
This parameter is given by the ratio of the particle terminal velocity 
to the atmospheric convective velocity, and depends on the particle size through the terminal velocity. 
It is commonly assumed that $f_{\rm sed}$ is constant throughout a cloud \citep{AM01,Morley13,Morley15},
but there is no guarantee that it must be for arbitrary convective clouds. 
The recent cloud model for Earth-like exoplanets by \citet{Zsom12} treats the microphysics of condensation, 
but greatly simplifies the coalescence processes by introducing the efficiency of precipitation as a free parameter.
The dust cloud model developed by \citet{Woitke03,Woitke04}, 
\citet{Helling06}, and \citet{Helling08}, which has recently been applied to clouds in hot Jupiters 
HD 209458b and HD 189733b \citep{Lee15,Helling16},
takes into account condensation and evaporation but not coalescence.
\citet{Woitke03} neglected coalescence because, according to \citet{Cooper03}, 
coalescence takes place much slower than condensation in brown dwarfs.
However, as noted by \citet{Woitke03}, coalescence becomes the only particle growth mechanism 
even in brown dwarfs if the supersaturation $s$, defined by the fractional excess of the partial pressure 
from saturation, is significantly low, e.g., $ s\ll 1\%$. 
While \citet{Cooper03} fixed the supersaturation to $1\%$, 
the actual supersaturation in an updraft depends on the number density of initial condensation nuclei, 
which is highly uncertain for exoplanets as well as for brown dwarfs.  

In this paper, we present a simple one-dimensional cloud model that is simple but 
takes into account the microphysics of coalescence as well as of condensation. 
We describe the basic equations and cloud microphysics in our model in Section 2. 
We test our model by comparing with the observations of the clouds on the Earth and Jupiter in Section 3.
We discuss the threshold velocity of sticking and the outlook on application to exoplanetary clouds in Section 4. 
We present a summary in Section 5.

\section{Model Description}
\subsection{Outline}\label{sec:outline}
\begin{figure}
\centering
\includegraphics[clip, width=\hsize]{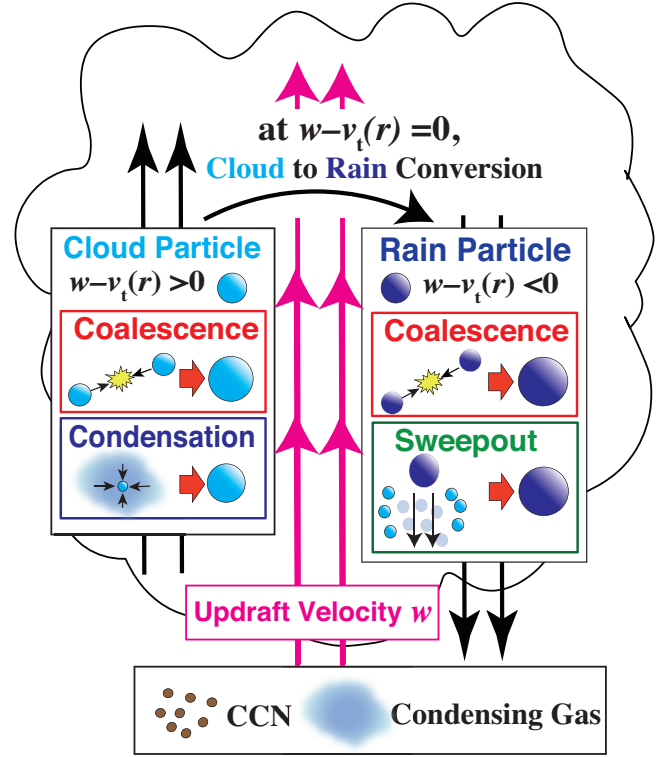}
\caption{Schematic illustration of our condensation--coalescence cloud model. We consider small ``cloud particles'' (light blue spheres)
and large ``rain particles'' (dark blue spheres) whose vertical velocity relative to the ground, $w-v_{\rm t}$, is positive and negative, respectively. 
The model includes particle growth due to condensation, coalescence, and sweepout (for the definitions of these processes, see Section~\ref{sec:outline}),
and also vertical transport due to gravitational settling and the updraft motion of the gas.}
\label{fig:model}
\end{figure}
Our model provides the vertical distributions of the mass and number densities of condensate particles
(Figure~\ref{fig:model}). 
We adopt a one-dimensional Eulerian framework in which the gas ascends at a vertical velocity $w$.
Each condensate particle is assumed to fall relative the the upwelling gas at a terminal speed $v_{\rm t}$, 
which is given by the balance between gas drag and gravity and is therefore a function of the particle radius 
$r$ (see Equation~\eqref{eq:vt_fit} for the expression of $v_{\rm t}$ adopted in this study). 
Thus, each particle has a net vertical velocity $w - v_{\rm t}(r)$. 
We divide the population of condensate particles into small ``cloud particles''  
whose net vertical motion is upward, $w-v_{\rm t}>0$, and large ``rain particles''
whose net vertical motion is downward, $w-v_{\rm t}<0$.
In this paper, we assume that initial cloud particles form through heterogeneous nucleation (see Section~\ref{sec:nuc} for details). 

We denote the total number density and mass density of the cloud (rain) 
particles by $N_{\rm c}$ ($N_{\rm r}$) and $\rho_{\rm c}$ ($\rho_{\rm r}$), respectively.
We assume that the cloud and rain particles have characteristic radii $r_{\rm c}$ and $r_{\rm r}$
and characteristic masses $m_c = (4\pi/3) \rho_{\rm int} r_c^3$ and $m_r = (4\pi/3) \rho_{\rm int} r_r^3$,
respectively, where $\rho_{\rm int}$ is the internal density of the particles. 
For liquid particles, $\rho_{\rm int}$ is equal to the material density of the condensate, $\rho_{\rm p}$,
while for solid particles, $\rho_{\rm int}$ can be lower than $\rho_{\rm p}$ 
because an aggregate of solid particles can be porous. 
The porosity of condensate particles can potentially affect 
the growth and motion of the particles as demonstrated by theoretical studies
on dust evolution in protoplanetary disks \citep[e.g.,][]{Ormel07,Okuzumi09}.
We here neglect this effect by assuming constant bulk density, $\rho_{\rm int} = \rho_{\rm s}$, 
but we plan to take it into account in future work. 

Our model determines the vertical distribution of $N_{\rm c}$, $\rho_{\rm c}$, $N_{\rm r}$, $\rho_{\rm r}$,
and the condensate vapor mass density $\rho_{\rm v}$
by numerically solving the set of vertically one-dimensional transport equations with terms 
representing condensation and coalescence (Sections~\ref{sec:transport} and \ref{sec:cond}--\ref{sec:sweep}).  
Condensation refers to particle growth through 
the accretion of supersaturated vapor, 
while coalescence refers to the growth through collisions with other condensate particles under gravity.  
In this study, we refer to the coalescence between cloud and rain particles as sweepout,
in order to distinguish the coalescence of two cloud particles or of two rain particles.\footnote{Sweepout is often termed ``accretion'' in the literature of cloud microphysical models 
\citep[e.g.,][]{Ziegler85}.} 
We neglect the condensation of vapor onto rain particles because the timescale 
of condensation growth is longer than the timescale of sedimentation for these large particles.
The characteristic masses and radii of two particle species (cloud and rain) 
are automatically determined by the mass and number densities 
via the relations $m_j = \rho_j/N_j$ and $r_j = (3m_j/4\pi \rho_{\rm p})^{1/3}$, respectively,
where $j = $ c for cloud particles and $j =$ r for rain particles.
Such a framework is called a double-moment bulk scheme in meteorology \citep[e.g.,][]{Ziegler85, Ferrier94}, and a characteristic size method in the planet formation community \citep[e.g.,][]{Birnstiel+12,Ormel14, Sato+16}. 
This characteristic size method allows us to simulate the growth of particles with much less computational time than that required with spectral bin schemes where the full size distribution of particles is evolved \citep{Birnstiel+12,Sato+16}. Our method is particularly useful for studying cloud formation over a wide parameter space.

Because the terminal velocity of particles generally increases as they grow,
there is a height $z = z_{\rm top}$ at which the net upward velocity 
of cloud particles reaches zero, i.e., $w  - [v_{\rm t}(r_{\rm c})]_{z=z_{\rm top}} = 0$. 
At this height, which we call the cloud top, we convert the cloud particles
into rain particles and allow them to fall and continue growing (see Figure~\ref{fig:model}). 
Our implementation of the cloud-to-rain conversion is described in Section~\ref{sec:conv}.

\subsection{Transport Equations}\label{sec:transport}
The transport equations used in our model are given by 
\begin{equation}
\frac{\partial N_{\rm c}}{\partial t}
= -\frac{\partial}{\partial z}[(w-v_{\rm t}(r_{\rm c}))N_{\rm c}]
- \left| \frac{\partial N_{\rm c}}{\partial t} \right|_{\rm coal}
-\left| \frac{\partial N_{\rm c}}{\partial t} \right|_{\rm sweep},
\label{eq:dNcdt}
\end{equation}
\begin{equation}
\frac{\partial \rho_{\rm c}}{\partial t}
= -\frac{\partial}{\partial z}[(w-v_{\rm t}(r_{\rm c}))\rho_{\rm c}]
+\left(\frac{\partial \rho_{\rm c}}{\partial t} \right)_{\rm cond}
-m_{\rm c}\left| \frac{\partial N_{\rm c}}{\partial t} \right|_{\rm sweep},
\label{eq:drhocdt}
\label{basic1}
\end{equation}
\begin{equation}
\frac{\partial N_{\rm r}}{\partial t}
= -\frac{\partial}{\partial z}[(w-v_{\rm t}(r_{\rm r}))N_{\rm r}] 
- \left| \frac{\partial N_{\rm r}}{\partial t} \right|_{\rm coal},
\label{eq:dNrdt}
\end{equation}
\begin{equation}							
\frac{\partial \rho_{\rm r}}{\partial t}
=  -\frac{\partial}{\partial z}[(w-v_{\rm t}(r_{\rm r}))\rho_{\rm r}] 
+m_{\rm c}\left| \frac{\partial N_{\rm c}}{\partial t} \right|_{\rm sweep},
\label{eq:drhordt}
\end{equation}
\begin{equation}
\frac{\partial \rho_{\rm v}}{\partial t}
= - \frac{\partial}{\partial z}(w\rho_{\rm v})- \left(\frac{\partial \rho_{\rm c}}{\partial t} \right)_{\rm cond},
\label{eq:drhovdt}
\end{equation}
where $(\partial \rho_c /\partial t)_{\rm cond}$ is the rate of increase in $\rho_c$ due to condensation,
$|\partial N_{\rm c}/\partial t|_{\rm coal}$ ($|\partial N_{\rm r}/\partial t|_{\rm coal}$) is the rate of 
decrease in $N_{\rm c}$ ($N_{\rm r}$) due to the coalescence of cloud (rain) particles themselves, 
and $|\partial N_{\rm c} /\partial t|_{\rm sweep}$ is the rate of decrease in $N_{\rm c}$ 
due to sweepout. 
The expressions for $(\partial \rho_c /\partial t)_{\rm cond}$, $|\partial N_j/\partial t|_{\rm coal}$,
and $|\partial N_{\rm c} /\partial t|_{\rm sweep}$ are given in Sections~\ref{sec:cond}--\ref{sec:sweep}. 

\subsection{Nucleation}\label{sec:nuc}
We assume that cloud particles form at the cloud base through 
the condensation of vapor onto small refractory grains that already exist in the atmosphere.
This process is known as heterogeneous nucleation, and such refractory grains are 
termed cloud condensation nuclei (CCN).  
On Earth CCN include sea salt, ash, and dust from the land \citep[e.g.,][]{Rogers89,Seinfeld06}.
Lacking information about CCN in other planets including exoplanets, 
we parametrize them with their number density $N_{\rm CCN}$ and radius $r_{\rm CCN}$.
The CCN number density is a particularly important parameter because 
it determines the number density and maximum reachable size of cloud particles growing through condensation. 

In principle, cloud particles may also form through homogeneous nucleation, where molecules in supersaturated vapor spontaneously collide to form initial nuclei. 
Although homogeneous nucleation is the simplest nucleation process, this occurs only at a supersaturation ratio much larger than unity because of the free energy barrier arising from the surface tension \citep[e.g.,][]{Rogers89,Marley13}. 
By contrast, heterogeneous nucleation generally occurs when the supersaturation ratio is slightly above unity
because the CCNs lower the free energy barrier \citep{Rogers89}.
However, if there are only a few CCNs available in the atmosphere, 
homogeneous nucleation would dominate over heterogeneous nucleation \citep{Woitke04}.
For simplicity, we ignore homogeneous nucleation in the present study, 
but we plan to include this effect in our future modeling. 

The cloud base is defined by the location above which the saturation vapor pressure $P_{\rm s}$ 
of condensing gas under consideration exceeds the partial pressure in the gas phase. 
This location is mainly determined by the vertical temperature profile of the atmosphere,
and weakly depends on the mixing ratio of the condensing gas under the cloud base.

\subsection{Condensation} \label{sec:cond}
If the cloud particle size $r_{\rm c}$ is much larger than the mean free path of condensing vapor molecules,  
the rate of increase in $\rho_{\rm c}$ due to condensation is given by \citep{Rogers89} 
\begin{equation}
\left( \frac{\partial \rho_{\rm c}}{\partial t}\right)_{\rm cond}
=  \frac{4\pi r_{\rm c}N_{\rm c}D(\rho_{\rm v}- \rho_{\rm s})}
{\left( \frac{L}{R_{\rm v}T}-1 \right)\frac{LD\rho_{\rm s}}{KT} +1 },
\label{eq:cond}
\end{equation}
where $L$ is the specific latent heat of vaporization, 
$N_{\rm c}$ is the number density of the cloud particles, 
$K$ is the coefficient of thermal conductivity of the atmosphere, 
and $D$, $R_{\rm v}$, and $\rho_{\rm s} = \rho_{\rm s}(T)$ are 
the diffusion coefficient, specific gas constant, and saturation vapor density of the vapor, respectively.
The saturation vapor density is related to the saturation vapor pressure $P_{\rm s}$ by
$\rho_{\rm s} = P_{\rm s}/R_{\rm v}T$. 
Equation~\eqref{eq:cond} neglects the effects of surface tension and nonvolatile solutes on
the saturation vapor pressure over the particles' surfaces.
This is a good approximation for activated condensate particles \citep{Rogers89}. 

The assumption that the particle sizes are larger than the atmospheric mean free path 
is justified for the clouds on the Earth and Jupiter we considered in Section~\ref{sec:tests}.  
The atmospheric mean free path in the terrestrial water clouds 
(where $T\approx 300~{\rm K}$ and $P \approx 1~{\rm bar}$) and in Jovian ammonia clouds
(where $T \approx 130$~{\rm K} and $P \approx 0.5~{\rm bar}$)
is $\sim 0.1\ {\rm \mu m}$, which is significantly smaller than the typical particle radius in the clouds
(see Figures \ref{fig:RICO} and \ref{fig:jupiter} in Section~\ref{sec:tests}), 
However, this assumption is not necessarily valid for clouds in exoplanets. 
For example, ZnS and KCl clouds on super-Earth GJ1214b could form at $P\sim 0.01\ {\rm bar}$ \citep{Morley13}
, where the mean free path is $\sim 100\ {\rm \mu m}$.
Furthermore, the observation of GJ1214b suggests the presence of high-altitude clouds at $P\sim {10}^{-5}\ {\rm bar}$  \citep{Kreidberg14,Morley15}, where the mean free path is as long as $l_{\rm g}\sim 1\ {\rm cm}$. 
When one applies our cloud model to such high-altitude clouds, one should replace 
Equation \eqref{eq:cond} by the expression in the free molecular regime \citep[for details, see ][]{Woitke03},
\begin{equation}
\left(\frac{\partial \rho_{\rm c}}{\partial t}\right)_{\rm cond}=4\pi r_{\rm c}^2 N_{\rm c}C_{\rm s}(\rho_{\rm v}-\rho_{\rm vs}),
\end{equation}
where $C_{\rm s}$ is the mean velocity of gas molecules.

\subsection{Coalescence} \label{sec:coal}
Under the assumption that the cloud and rain particle size distributions are narrow,
the rate of decrease in $N_j$ ($j = $ c or r') due to coalescence
is approximately given by
 
\begin{eqnarray}
\nonumber
\left| \frac{\partial N_j}{\partial t} \right|_{\rm coal}
&\approx& \frac{1}{2}\pi (r_j+r_j)^2 N_j^2 \Delta v(r_j) E\\
&=& 2\pi r_j^2 N_j^2 \Delta v(r_j) E,
\label{eq:coal}
\end{eqnarray}
where $\Delta v$ is the relative velocity between the particles induced by gravitational settling, 
$E$ is the collection efficiency defined by the ratio of the effective collisional cross section
to the geometric one (see below). 
The factor $1/2$ prevents double counting of the collisions. 
We have used that the geometric collisional cross section 
between two similar-sized particles is approximately given by $\pi (r_j+r_j)^2$. 
We express the differential settling velocity 
as $\Delta v = \epsilon v_{\rm t}(r_j)$, where the factor $\epsilon (<1)$ 
encapsulates the effect of non-zero particle size dispersion.  
\citet{Sato+16} and \citet{Krijt16} show that 
when the differential drift velocity is proportional to the particle size,
bulk schemes with $\epsilon \approx 0.5$ 
best reproduce the results of spectral bin schemes that take into account the full size distribution.
In this study, we adopt $\epsilon = 0.5$ for arbitrary values of $r_{j}$. 

The collection efficiency $E$ accounts for the effect of the gas flow around a large particle 
moving relative to the background gas: the gas flow sweeps aside particles 
that are aerodynamically well coupled to the gas (see, e.g., \citealt{Slinn74}, \citealt{Pru97}, their Chapter 14). 
For a collision between two particles of radii $r$ and $r'$ ($r>r'$), 
the collection efficiency can be expressed in terms of the Stokes number 
\begin{equation}
{\rm Stk}= \frac{v_{\rm t}(r')|v_{\rm t}(r)-v_{\rm t}(r')|}{gr},
\label{eq:Stk}
\end{equation} 
which is approximately the ratio between the stopping time $= v_{\rm t}(r')/g$
and crossing time $\sim r/|v_{\rm t}(r)-v_{\rm t}(r')|$ of the smaller particle.
When ${\rm Stk} \ll 1$, the smaller particle is strongly coupled to the flow around the large particle.
To zeroth order, $E$ behaves as $E \approx 0$ at ${\rm Stk} \ll 1$ and 
as $E\approx 1$ at ${\rm Stk} \gg 1$ \citep{Rossow78}. 
In this study, we evaluate $E$ using a smoother analytic function \citep[][their Equation~(99)]{Guillot14}. 
\begin{equation}
E={\rm max}[0\ ,1-0.42\ {\rm Stk}^{-0.75}],
\label{eq:E}
\end{equation}
which vanishes at ${\rm Stk}\la 0.3$ and approaches unity at ${\rm Stk} \gg 1$
(see Figure 12 of \citealt{Guillot14}).
This expression assumes that flow around the particle is laminar.  
If the gas flow is turbulent, the collection efficiency can be higher than assumed here \citep{Homann+16}. 
We approximate ${\rm Stk}$ in $E$ as ${\rm Stk} \approx v_{\rm t}(r_j) \epsilon v_j/(gr_j)$. 
Because ${\rm Stk}$ generally increases with $r_j$, coalescence 
occurs only after the particle size exceeds a threshold above which ${\rm Stk}>1$.
Therefore, the production of precipitating rain droplets through coalescence 
requires growth beyond this threshold by condensation \citep{Pru97}.

Equation~\eqref{eq:E} applies when the background gas behaves as a fluid, i.e., 
the particle radius is much larger than the mean free path of the gas molecules. 
In the opposite case where the gas behaves as a free molecular flow, one may assume $E = 1$. 
The free-molecular flow regime is expected to govern the collisional growth of particles at high altitudes where the gas density is low.


\subsection{Sweepout}  \label{sec:sweep}
Similar to Equation~\eqref{eq:coal}, the rate of decrease in $N_{\rm c}$ due to sweepout is given by 
\begin{equation}
\left| \frac{dN_{\rm c}}{dt} \right|_{\rm sweep}
=\pi (r_{\rm r}+r_{\rm c})^2 |v_{\rm t}(r_{\rm r})-v_{\rm t}(r_{\rm c})|N_{\rm r}N_{\rm c}E.
\end{equation}
where the collection efficiency $E$ is given by Equation~\eqref{eq:E}
with ${\rm Stk}=v_{\rm t}(r_c)|v_{\rm t}(r_{\rm r})-v_{\rm t}(r_{\rm c})|/(gr_{\rm r})$ (see Equation~\eqref{eq:Stk}).

\subsection{Cloud-to-Rain Conversion}\label{sec:conv}
Cloud-to-rain conversion occurs at the height $z=z_{\rm top}$
where the terminal velocity of cloud particles $v_{\rm t}(r_{\rm c})$ equals the updraft velocity $w$. 
Once the terminal velocity of cloud particles exceeds the updraft velocity,
we numerically fix the net vertical velocity, $w-v_{\rm t}(r_{\rm c})$, to zero, 
and let the cloud particles evolve into rain particles 
at the rate given by $t_{\rm conv}^{-1} = \beta(t_{\rm cond}^{-1} + t_{\rm coal}^{-1})$, 
where $t_{\rm cond}^{-1} \equiv (\partial \rho_{\rm c}/\partial t)/\rho_{\rm c}$ and 
$t_{\rm coal}^{-1} \equiv |\partial N_{\rm c}/\partial t|/N_{\rm c}$ are the rates of growth due to condensation and coalescence,
respectively, and $\beta$ is a numerical factor (see also below). 
The equations that describe the cloud-to-rain conversion are then given by 
\begin{equation}
\left|\frac{\partial N_{\rm c}}{\partial t}\right|_{z=z_{\rm top}}
= -\frac{N_{\rm c}}{t_{\rm conv}},
\label{eq:Nc_top}
\end{equation}
\begin{equation}
\left|\frac{\partial \rho_{\rm c}}{\partial t}\right|_{z=z_{\rm top}}
= -\frac{\rho_{\rm c}}{t_{\rm conv}},
\label{eq:rhoc_top}
\end{equation}
\begin{equation}
\left|\frac{\partial N_{\rm r}}{\partial t}\right|_{z=z_{\rm top}}
= \frac{N_{\rm c}}{t_{\rm conv}},
\label{eq:Nr_top}
\end{equation}
\begin{equation}
\left|\frac{\partial \rho_{\rm r}}{\partial t}\right|_{z=z_{\rm top}}
= \frac{\rho_{\rm c}}{t_{\rm conv}}.
\label{eq:rhor_top}
\end{equation}

Strictly speaking, our cloud--rain two-population model breaks down near the cloud top, 
where the true size distribution of condensate particles (which is not resolved in our model) 
cannot be approximated with two peaks that are well separated from each other. 
The numerical factor $\beta$ we introduced above arises from the lack of information
about the continuous particle size distribution from cloud to rain particles.
We find that $\beta \ga 1$--$100$ causes an oscillating motion of the cloud top that 
prevents us from obtaining a steady solution. By contrast, $\beta \la 1$ provides 
a stable steady solution, and we find that the vertical distribution of cloud and rain 
well below the cloud top is insensitive to the choice of $\beta$ as long as $\beta \la 1$ (see Section \ref{sec:Earth},\ref{sec:Jupiter}).
We will adopt $\beta = 0.1$ in the test simulations presented in the following section.

\subsection{Boundary Conditions}
The lower boundary of the computational domain is set to the cloud base,
which refers to the height above which
the saturation ratio $S \equiv \rho_{\rm v}/\rho_{\rm s}$ exceeds unity. 
To determine the location of the could base, 
we introduce the mixing ratio of the cloud-forming vapor below the cloud base as an input parameter.
At the cloud base, we set the number density and radius of cloud particles to be equal to   
those of CCN, $N_{\rm CCN}$ and $r_{\rm CCN}$, respectively. 
For rain particles, we simply allow them to leave the computational domain 
with the (downward) flux determined just above the cloud base. 

The upper boundary is taken to be the cloud top, and the boundary conditions 
at that location are given by Equations~\eqref{eq:Nc_top}--\eqref{eq:rhor_top}.
Neither cloud particles nor rain particles are allowed to go above the cloud top because,
by definition, the net vertical velocities of the particles vanish at the cloud top. 

\section{Test Applications} \label{sec:tests}
Now we test our cloud model by comparing the observational data 
of water clouds on the Earth and of ammonia ice clouds on Jupiter.
For simplicity, we assume that the vertical structure and updraft motion of 
the background atmosphere is independent of the presence of clouds. 
We numerically solve Equations~\eqref{eq:dNcdt}--\eqref{eq:drhovdt} 
with Equations~\eqref{eq:Nc_top}--\eqref{eq:rhor_top}, 
under the fixed boundary conditions at the cloud base,
until steady profiles of $N_{\rm c}$, $\rho_{\rm c}$, $N_{\rm r}$, 
and $\rho_{\rm r}$ are obtained. 
To do so, we discretize the vertical coordinate $z$ 
into linearly spaced bins of width $\Delta z$ ($10~{\rm m}$ for terrestrial water clouds and $20~{\rm m}$ for Jovian ammonia clouds), and integrate the equations in time using the first-order explicit scheme.

\subsection{Water Clouds on the Earth}\label{sec:Earth}
We focus on trade-wind cumuli, which are relatively shallow water clouds 
but are yet deep enough to develop precipitation. 
\citet{van11} conducted comparative studies of 
terrestrial cloud models using the data of trade-wind cumuli in the northwestern Atlantic ocean
obtained from the Rain in Cumulus over the Ocean (RICO) Field Campaign \citep{Rauber07}.  
Following \citet{van11}, we examine how accurately our cloud model 
reproduces the vertical cloud distribution from the RICO campaign. 

Following \citet{WL73}, we construct the vertical temperature profile 
by using the dry adiabatic lapse rate ($g/c_{\rm p}=9.8~{\rm K~km^{-1}}$, 
where $c_{\rm p}$ is the specific heat at constant pressure) below the cloud base 
and the wet adiabatic lapse rate \citep[e.g., the Equation (7) of][]{Atreya05} above the cloud base.
The vertical pressure profile is calculated by assuming hydrostatic equilibrium 
(which is a good approximation as long as the updraft motion is slow).
In accordance with \citet{van11}, we fix the surface temperature and 
CCN number density to $298\ {\rm K}$ and $70\ {\rm cm}^{-3}$, respectively,
and adjust the water vapor mixing ratio (or the specific humidity) below the cloud base 
so that the cloud base is located at $500\ {\rm m}$ from the ground. 
The updraft velocity is taken to be either $0.9~{\rm m~s^{-1}}$ or $2.0~{\rm m~s^{-1}}$,
which represent the typical updraft velocities in the dense and diffuse parts of the clouds,
respectively, in the simulations by \citet[][their Figure 5]{van11}.

We approximate the vapor pressure for liquid water with the Arrhenius function 
\begin{equation}
p_{\rm s,H_2O}=611{\rm exp}\left[\frac{L}{R_{\rm v}}\left( \frac{1}{273~{\rm K}}-\frac{1}{T} \right) \right]~{\rm Pa} ,
\end{equation}
where the temperature is in K and $L=2.5\times {10}^6\ {\rm J\ {kg}^{-1}}$ \citep{Rogers89}.
The thermal conductivity $K$ and dynamic viscosity $\eta$ of air are set to 
$2.4\times{10}^{-2}\ {\rm W\ m^{-1}\ K^{-1}}$ and $1.7\times{10}^{-5}\ {\rm Pa\ s}$,
respectively, and the diffusion coefficient $D$ of water vapor in air is set to 
$2.2\times{10}^{-5}\ {\rm m^2\ s^{-1}}$.
These are the values at $273.0\ {\rm K}$ and $100\ {\rm kPa}$ \citep{Rogers89}.
The CCN radius $r_{\rm CCN}$ is assumed to be $0.5~\micron$; the results presented below are insensitive to the choice 
of $r_{\rm CCN}$ as long as it is taken to be smaller than $1~\micron$.

\begin{figure*}
\centering
\includegraphics[clip,width=\hsize]{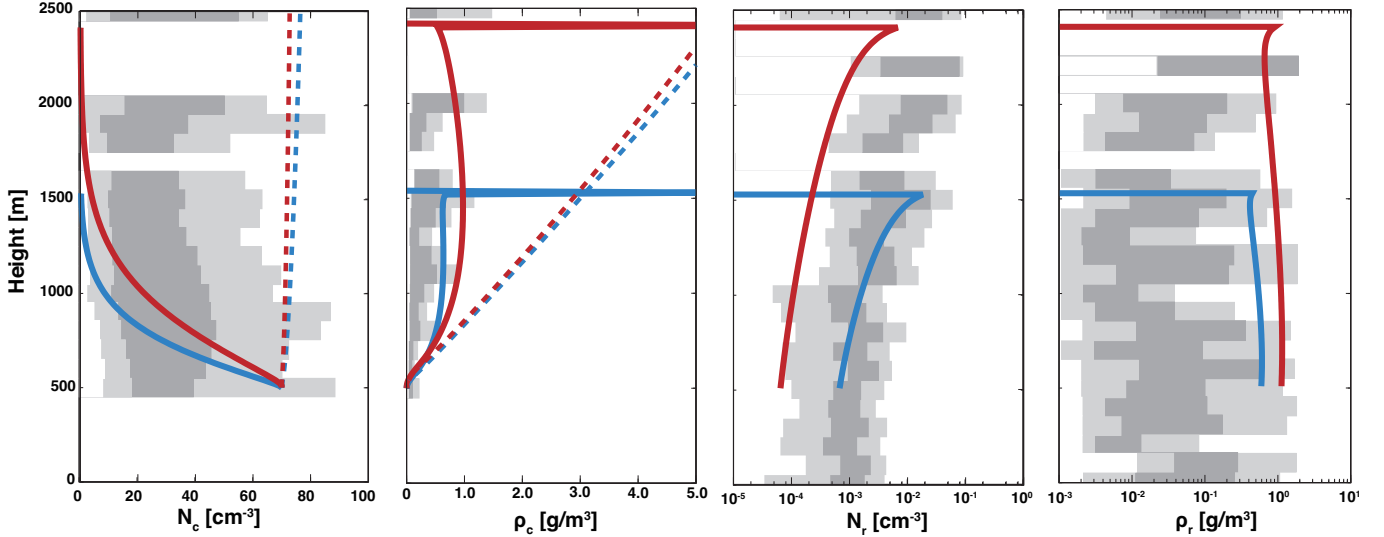}
\caption{
\label{earth1}
Vertical structure of trade cumulus clouds on the Earth
derived from our model calculations as well as from the RICO observations.
The four panels show, from left to right, the number and mass densities of 
cloud particles, $N_{\rm c}$, $\rho_{\rm c}$, and those of rain particles, 
$N_{\rm r}$, and $\rho_{\rm r}$, as a function of the height $z$ from the ground. 
The blue and red solid lines show the steady-state distributions obtained from 
our condensation--coalescence model for $w=0.9$ and $2.0~{\rm m~s^{-1}}$, respectively. 
The dotted lines show the results from the model neglecting coalescence.
The light and dark gray-shaded areas span the $5$ to $95\%$ and $25$ to $75\%$ ranges of the RICO observation data, respectively, taken from Figure 8 of \citet{van11}.
}
\end{figure*}
\begin{figure}
\centering
\includegraphics[clip,width=\hsize]{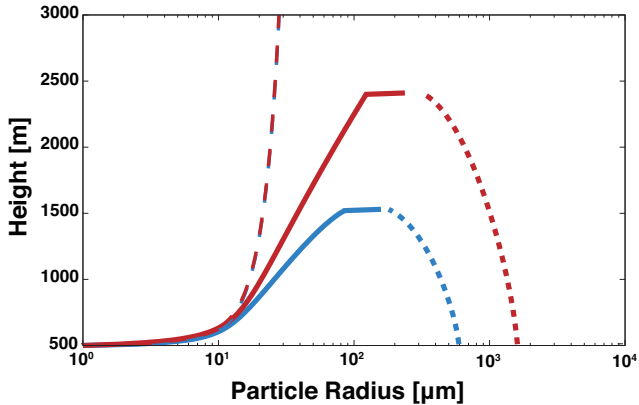}
\caption{
\label{fig:RICO}
Vertical distributions of the cloud particle radius $r_{\rm c}$ (solid lines) and 
rain particle radius $r_{\rm c}$ (dotted line) from our condensation--coalescence cloud calculations 
shown in Figure~\ref{earth1}. The blue and red lines show the results for $w=0.9$ and $2.0~{\rm m~s^{-1}}$, respectively. The dashed lines show the vertical distribution of the cloud particle radius neglecting coalescence.
}
\end{figure}

Figures~\ref{earth1} and \ref{fig:RICO} present the results of our test calculations.
In Figures~\ref{earth1}, we show the steady-state vertical profiles of the number and mass densities 
of the cloud and rain particles, for $w = 0.9$ and $2.0~{\rm m~s^{-1}}$. 
To highlight the effects of coalescence, we also show  
the results of simulations without the coagulation and sweepout terms (see the dotted lines).
The sizes of the cloud and rain particles as a function of the height $z$ from the ground 
are shown in Figure~\ref{fig:RICO}. In these simulations, the size of the cloud particles 
reach $\sim 100~{\rm \micron}$ at $z \approx 1500~{\rm m}$ for $w = 0.9~{\rm m~s^{-1}}$ 
and at $z \approx 2200~{\rm m}$ for $w = 2.0~{\rm m~s^{-1}}$.
At these heights, the net vertical velocity $w-v_{\rm t}(r_{\rm c})$ of the cloud particles becomes zero, 
and we convert them into rain particles that are allowed to continue growing as they fall toward the ground.

Now we compare our simulation results with the RICO flight observations.
The light and dark gray-shaded areas in Figure~\ref{earth1} indicate the $5$--$95\%$ 
and  $25$--$75\%$ ranges of the observed values, respectively, taken from Figure 8 of \citet{van11}.
Overall, we find that our condensation--coalescence model reproduces 
the observations to order of magnitude. 
The predicted values of the cloud and rain densities 
fall within the $5$--$95\%$ range of the observations, except at high altitudes 
$z \ga 1700~{\rm m}$ where the result for $w = 0.9~{\rm m~s^{-1}}$ considerably 
underestimates the cloud number density. 
By contrast, the model neglecting coalescence is found to overestimate 
the mass density of cloud particles by up to an order of magnitude.
Furthermore, this condensation-only model fails to reproduce precipitation 
because the maximum particle size reachable with condensation is too small to fall against 
an updraft of $w \sim 1~{\rm m~s^{-1}}$ (see the dashed lines in Figure \ref{fig:RICO}). 
With coalescence, cloud particles do grow large enough to  
start falling as rain particles as we already described above.

Although coalescence resolves the order-of-magnitude 
discrepancy between the model predictions and observations, 
the cloud and rain mass densities are still systematically higher than the averages of the observed values. 
This indicates that the updrafts that produce the observed clouds entrain dry ambient air \citep[e.g.,][]{Pru97}.
Entrainment reduces the temperature and humidity the updraft, both of which act to
suppress the condensation growth of cloud particles. 
The suppressed growth in turn leads to a slower decrease in the cloud number density with height, 
because coalescence takes place only after the particles grow sufficiently large (see Section~\ref{sec:coal}).
Therefore, a model including entrainment might better reproduce 
both the number and mass density of cloud particles.
However, the modeling of entrainment within a 1D framework necessarily introduces 
additional poorly constrained free parameters \citep[see e.g.,][Chapter 12.7, 12.8]{Pru97}, which we avoid in this study.

\begin{figure}
\centering
\includegraphics[clip,width=\hsize]{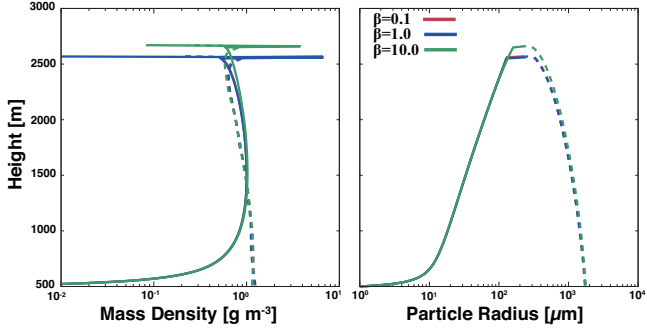}
\caption{
\label{fig:beta_earth}
Influences of numerical factor $\beta$ on the vertical distributions of the cloud particle radius $r_{\rm c}$ and mass density $\rho_{\rm c}$ (solid lines), and the rain particle radius $r_{\rm r}$ and mass density $\rho_{\rm r}$ (dashed line). 
The red, blue, and green lines show the simulation results for $\beta=0.1$, $1.0$, and $10.0$, respectively.
The CCN number densities and updraft velocity are set to $N_{\rm CCN}=100~{\rm {cm}^{-3}}$ and $w=2.0~{\rm m/s}$.
}
\end{figure}

Figures \ref{earth1} and \ref{fig:RICO} show that 
there is a jump in the cloud particle radius and a peak in the cloud mass density at the cloud top.
This is an artifact arising from our treatment of the cloud--rain conversion at this location. 
As emphasized in Section~\ref{sec:conv}, our cloud--rain two-population model does not 
perfectly treat the cloud top where the two populations in reality merge into a single distribution of particles. 
In fact, we find that the jump and peak features in $r_{\rm c}$ and $\rho_{\rm c}$, 
as well as the height of the cloud top, weakly depends on 
the value of the numerical factor $\beta$ introduced in our cloud--rain conversion calculations.
This is shown in Figure \ref{fig:beta_earth}, where we plot  the vertical distributions 
of the sizes and mass densities of cloud and rain particles for $\beta=0.1$, $1.0$, and $10.0$.
However, we confirm that this artifact has little effect on the vertical distributions well below the cloud top.


\subsection{Ammonia Ice Clouds on Jupiter}\label{sec:Jupiter}

\begin{figure}
\centering
\includegraphics[clip,width=\hsize]{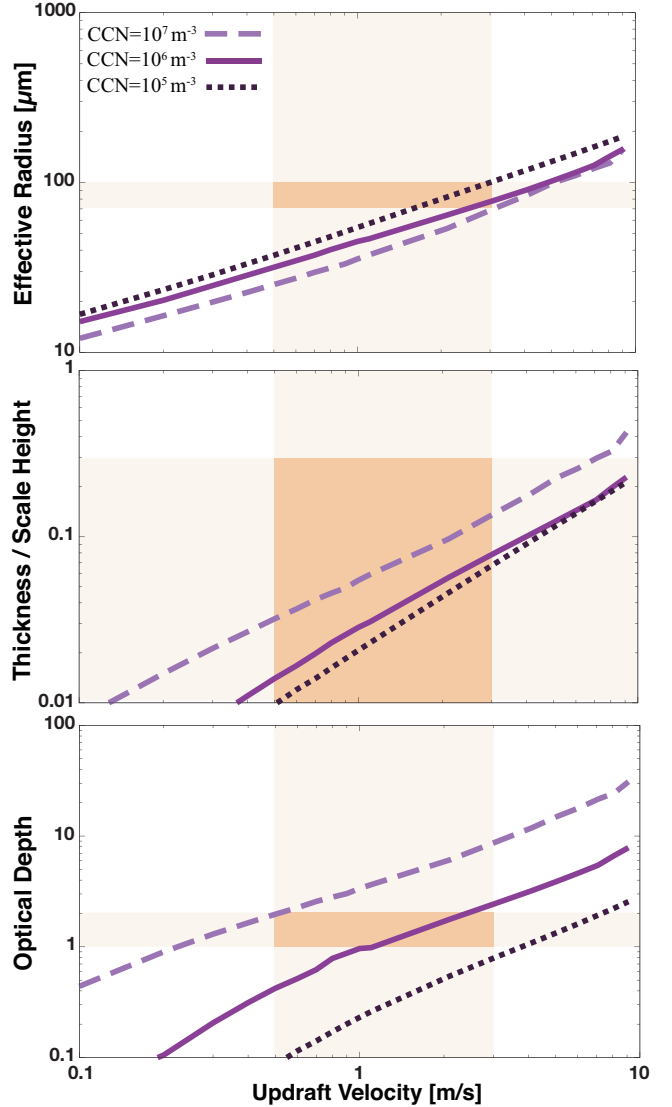}
\caption{Effective radius $r_{\rm eff}$ (top panel), cloud geometric thickness $H_{\rm p}$ normalized by pressure scale height $H_{\rm g}$, (middle panel), 
and cloud optical depth $\tau$ at visible wavelengths (bottom panel) for Jovian ammonia clouds. 
The solid lines show the predictions from our condensation--coalescence model 
for different values of the updraft velocity $w$ ($x$-axis) and CCN density $N_{\rm CCN}$ (dashed lines for ${10}^7~{\rm m^{-3}}$, solid lines for ${10}^6\ {\rm m^{-3}}$, and dotted lines for ${10}^5\ {\rm m^{-3}}$). 
The orange horizontal bars indicate the retrievals from the Voyager IRIS observations by \citet{Carlson94}, while  
the vertical bars indicate the updraft velocity inferred from the 2D simulations of Jovian moist convection by \citet[][their Section 3.1 and Figure 9]{Sugiyama14}. 
The prediction from the condensation--coalescence model satisfies all these constraints when $w$ and $N_{\rm CCN}$ are taken to be
$2$ -- $3~{\rm m~s^{-1}}$ and $N_{\rm CCN} \approx 10^{6}~{\rm m^{-3}}$, respectively. 
}
\label{fig:jupiter}
\end{figure}
We also attempt to reproduce the observations of ammonia clouds on Jupiter.
Following \citet{AM01}, we focus on ammonia ice clouds that cover Jupter's upper troposphere where $P \sim 0.7~{\rm bar}$ 
and $T \sim 130$--$140~{\rm K}$ \cite[e.g.,][]{West86}.
In accordance with the measurements by the Galileo probe, we model the vertical temperature profile as $T= 166~{\rm K} + \Gamma(z-z_0)$,
where $\Gamma = -2~{\rm K~km^{-1}}$ is the lapse rate and $z_0$ is the height at which $P = 1.0~{\rm bar}$ \citep{Seiff98}, 
and set the mixing ratio of ammonia gas under the cloud base to be $6.64\times{10}^{-4}\ {\rm kg~{kg}^{-1}}$ \citep{Wong04}.
The vertical pressure profile is determined under the assumption of hydrostatic equilibrium. 
We take the updraft velocity and CCN number density as free parameters ranging from $0.1\ {\rm m~s^{-1}}$ to $10\ {\rm m~s^{-1}}$ 
and from ${10}^3\ {\rm m^{-3}}$ to ${10}^8\ {\rm m^{-3}}$, respectively.
For the the vapor pressure of ammonia ice, we use the expression by \citet[][see also \citealt{AM13}]{AM01},
\begin{equation}
p_{\rm s,NH_3}={\rm exp}\left(10.53-\frac{2161.0}{T}-\frac{86596.0}{T^2}\right)\ {\rm bar},
\end{equation}
where the temperature is in K.
The thermal conductivity of the atmosphere is taken to be $K= 9.0\times 10^{-2}\ {\rm W\ m^{-1}\ K^{-1}}$, 
which is the value for Jovian atmosphere at $T=134.3\ {\rm K}$ \citep{Hansen79}.
The dynamic viscosity of the atmosphere is given by $\eta = 6.7\times{10}^{-6}\ {\rm Pa\ s}$ 
based on the formula for the mixture gas of ${\rm H_2}$ and ${\rm He}$ \citep{Woitke03} together 
with the assumption $T \approx 130\ {\rm K}$.
The diffusion coefficient of ammonia gas is given by the formula $D= {2\eta}/({3\rho}f)$, where
$\rho$ is the atmospheric density and $f=5$ for ammonia vapor (Equation~(14) of \citealt{Rossow78}).
Following \citet{AM01}, we take the bulk density of ammonia ice $\rho_{\rm p}$ to be $0.84~{\rm g~cm^{-3}}$.
As in the previous subsection, the CCN radius $r_{\rm CCN}$ is assumed to be $0.5~\micron$.

We compare the steady-state vertical profiles of the cloud and rain particles obtained from our model
with the retrievals by \citet{Carlson94} based on the infrared observations of Jovian ammonia clouds
in the Equatorial Zone and North Tropical Zone from Voyager instrument IRIS. 
\citet{Carlson94} retrieved the effective particle radius $r_{\rm eff}$, geometric thickness $H_{\rm p}$, and optical depth $\tau$ at the wavelength of $0.5~\micron$,
where the effective radius refers to the area-weighted average of the radius of visible condensate particles \citep[see, e.g.,][]{Kokhanovsky04}. 
We evaluate the effective radius in our simulated clouds as 
\begin{equation}
r_{\rm eff}=\frac{\int (r_{\rm c}^3N_{\rm c}+r_{\rm r}^3N_{\rm r}) \exp(-\tau_z)dz}{\int (r_{\rm c}^2N_{\rm c}+r_{\rm r}^2N_{\rm r}) \exp(-\tau_z)dz}.
\end{equation}
where $\tau_z$ is the optical thickness above height $z$. 
The factor $\exp(-\tau_z)$ accounts for the fact that one can only observe particles residing at $\tau_z \la 1$.  
In the calculation of $\tau$ and $\tau_z$, we apply the geometric optics approximation to the extinction cross section at visible wavelengths. 
The geometric thickness of the simulated cloud is taken to be the distance between the cloud base and cloud top in this study. 
\begin{figure}
\centering
\includegraphics[clip,width=\hsize]{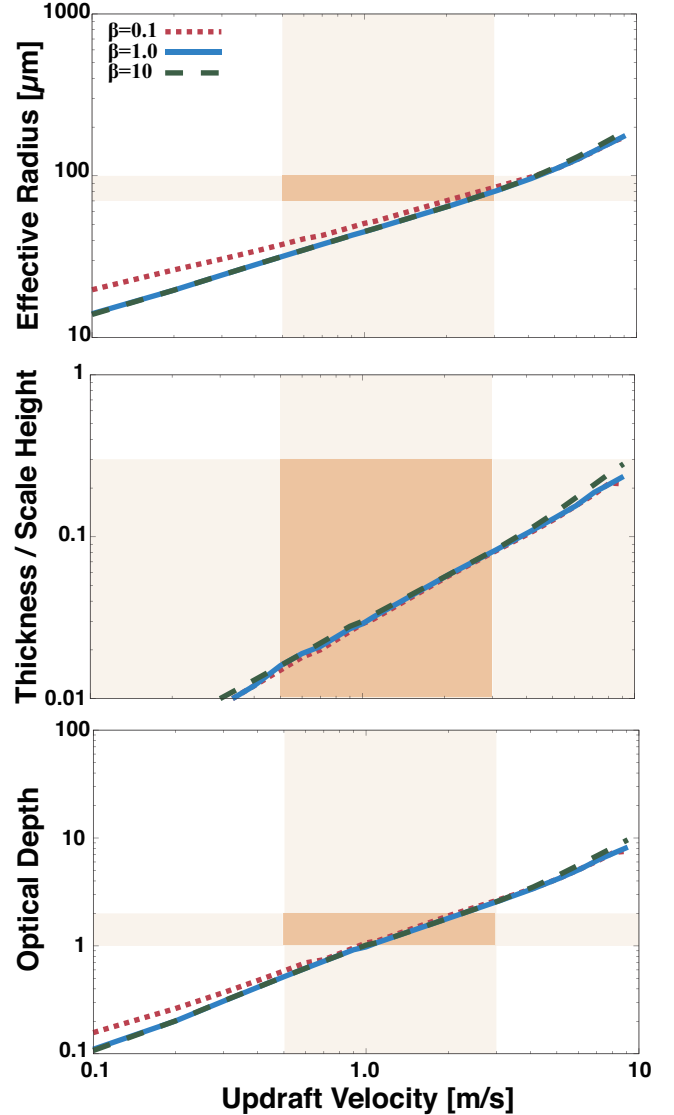}
\caption{Influence of $\beta$ on the results for Jovian ammonia ice clouds.
The value of $N_{\rm CCN}$ is fixed to ${10}^6~{\rm m^{-3}}$.
The dotted red lines show the simulations for $\beta=0.1$, the solid blue lines show 
for $\beta=1.0$, and the dashed dark-green lines show for $\beta=10.0$, respectively.
}
\label{fig:beta_jup}
\end{figure}

The results are summarized in Figure~\ref{fig:jupiter}.
Here, the solid lines show the values of $r_{\rm eff}$, $H_{\rm p}$, and $\tau$ 
from our condensation--coalescence model for different sets of the updraft velocity $w$ and CCN density $N_{\rm CCN}$.
The orange horizontal bars indicate the retrievals by \citet[][]{Carlson94}: 
$r_{\rm eff} =70$--$100~{\rm \micron}$, $H_{\rm p} \leq 0.3H_{\rm g}$, and $\tau = 1.2$--$2.0$,
where $H_{\rm g}=20~{\rm km}$ is the pressure scale height and the range for $r_{\rm eff}$ is based on the interpretation by \citet{AM01}.  
We find that the predictions from the condensation--coalescence model satisfies all these observational constraints 
when the updraft velocity $w$ and CCN number density $N_{\rm CCN}$ 
are assumed to be $\approx2$--$3~{\rm m~s^{-1}}$
and $\approx {10}^6 \ {\rm m}^{-3}$ (see the solid black lines), respectively. 
If $N_{\rm CCN} \ga {10}^7 \ {\rm m}^{-3}$ (see the dashed lines), 
the predicted optical depth falls within the retrieved range only when $w \approx 0.2$--$0.5~{\rm m~s^{-1}}$; however, for this range of $w$, the predicted effective radius is too small to be consistent with the retrieval.
If $N_{\rm CCN} \la {10}^5~{\rm m^{-3}}$ (see the dotted lines),
our prediction reproduces the retrieved optical depth 
only when $w \approx 3$--$7~{\rm m~s^{-1}}$, but then overestimates the effective radius from the retrieval.

The results presented in Figure~\ref{fig:jupiter} are little affected by the choice of the $\beta$ parameter introduced in our cloud-top treatment, as shown in Figure~\ref{fig:beta_jup}. 
One can see that the predictions converge for $\beta \ga 1$.
For $\beta = 0.1$, the predicted effective radius and optical depth at $w \ga 0.5~{\rm m/s}$ 
are higher than the converged values, 
but the deviation from the converged value is as small as a factor of less than 2.

We compare the updraft velocity of $2$--$3\ {\rm m~s^{-1}}$
with other calculations to validate our best fit value.
First, the mixing length theory formulated by \citep{AM01} suggested the updraft velocity at ammonia cloud region is approximately $1$--$3~{\rm m~s^{-1}}$,
which corresponds to the eddy diffusion coefficient $K_{\rm z}=2$--$6 \times {10}^8~{\rm {cm}^2 ~s^{-1}}$ 
and the mixing length equal to $H_{\rm g}=20~{\rm km}$.
Furthermore, the 2D simulations of the Jovian moist convection show that 
the updraft velocity at ammonia cloud region is $0.5$--$3\ {\rm m~s^{-1}}$ \citep{Sugiyama14}
shown in the vertical orange bar of Figure \ref{fig:jupiter}. 
Therefore, our best fit value of updraft velocity of $2$ -- $3\ {\rm m~s^{-1}}$ is 
a realistic value for the Jovian ammonia cloud region.

Unfortunately, there is no direct observational constraint 
on the CCN number density  for Jovian ammonia clouds.
However, an important hint for the CCN density can be obtained from 
the in-situ observation of the Jovian atmosphere in a
relatively particle-free hot spot by the Galileo probe. 
The observation 
showed that a concentration of small particles 
of a mean radius of $0.5$--$5\ {\rm \micron}$ and a number density of
$1.9\times{10}^5$--$7.5\times{10}^6\ {\rm m^{-3}}$ was present 
at a height where ammonia clouds form \citep{Ragent98}.
It is unlikely that the observed particles had experienced coalescence, 
because the particles are too small to collide with
each other (i.e., for particles smaller than a micron in radius, 
${\rm Stk}<0.5$ and hence $E<0.3$ in the ammonia cloud forming region).
Therefore, we can infer that the number density of CCN, which is approximately equal to
the number density of small particles before coalescence sets in, 
was $\sim {10}^5$--${10}^7~{\rm m^{-3}}$ in the hot spot the Galileo probe entered.
Interestingly, this value is consistent with our prediction for the CCN density 
in the Equatorial Zone and North Tropical Zone (see above). 
We do not think that this comparison justifies our cloud model, 
because the CCN density in relatively cloud-free hot spots is not necessarily equal 
to that in cloudy Equatorial Zone and North Tropical Zone.  
Rather, this comparison {\it suggests} that the CCN concentrations in the two regions could be similar.

\section{Summary and Outlook}
We have developed a new cloud model for exoplanets and brown dwarfs
that is simple but takes into account 
the microphysics of both condensation and coalescence. 
Our model produces the vertical distributions of the mass and number densities of cloud and rain particles
as a function of physical parameters, including the updraft velocity, the mixing ratio of the condensing gas at the cloud base, and the number density of CCN.
Therefore, our model will be useful to understand how 
the dynamics, compositions, and nucleation processes in exoplanetary atmospheres
would affect the vertical structure of exoplanetary clouds via cloud microphysics.

We have tested our model by comparing with the observations of the terrestrial water clouds and the Jovian ammonia clouds. 
For terrestrial water clouds, our model plausibly reproduces the observed vertical distributions of the cloud mass and number densities from in situ observations 
when we assume the terrestrial typical updraft velocity, height of cloud base, and CCN number density. 
For Jovian ammonia clouds, our model simultaneously reproduces the cloud optical depth, the geometric thickness, and the particle effective radius in the Equatorial Zone and North Tropical Zone retrieved from Voyager measurements 
when we assume the updraft velocity of $w\approx2$--$3\ {\rm m~s^{-1}}$ and the CCN number density 
of $N_{\rm CCN}\approx {10}^{6}\ {\rm m}^{-3}$. 
Our best-fit updraft velocity is consistent with estimates from mixing theory and 
from cloud convection simulations. The best-fit CCN density is close to 
the number density of small particles in a hot spot measured by the Galileo probe, suggesting 
that the CCN density in the Equatorial Zone and North Tropical Zone is similar to that in hot spots. 
The good agreement between our predictions and the observations indicates that 
the coalescence of condensate particles is an important process of cloud formation, 
not only in terrestrial water clouds but also in Jovian ice clouds.

Equation~\eqref{eq:coal} assumes that two particles stick whenever they collide. 
However, this assumption breaks down if the collision velocity is so high 
that the collision results in bouncing or fragmentation of the particles. 
In principle, solid particles are less sticky than liquid particles because 
a harder particle has a smaller contact area and hence a small binding energy 
associated with intermolecular forces \citep{Rossow78}.
Our future modeling will take into account the potentially low sticking efficiency of solid dust particles.

We have also assumed that the internal density of condensate particles is constant.  
However, the internal density can decrease as the particles grow into porous aggregates 
through coalescence \citep[e.g.,][]{Blum00}. 
Because porous aggregates have a larger aerodynamical cross section than compact particles of the same mass,
porous aggregates might ascend to higher altitudes than compact ones
and provide featureless transmission spectra of exoplanetary atmospheres. 
On the other hand, the coalescence of porous aggregates can be faster than that of compact particles \citep{Okuzumi+12}. If this is the case in exoplanetary atmospheres, 
the porosity evolution might prevent the formation of high-altitude clouds.
Recent theoretical studies of the grain growth in protoplanetary disks have yielded a detailed model for the porosity evolution of grain aggregates based on grain contact mechanics (e.g., \citealt{Kataoka13}). 
We will study the impact
of porosity evolution on cloud formation by using these theories in future studies. 

\acknowledgments 
The authors thank the anonymous {referee} for useful comments. 
We thank {Hanii} Takahashi and {Yuka} Fujii for useful discussions about {observations} of terrestrial clouds, and {Chris} Ormel for many helpful comments on the manuscript of this paper, and {Neal} Turner, {Shigeru} Ida, {Masahiro} Ikoma, {Mark} Marley, {Yasuto} Takahashi, and {Xi} Zhang for useful comments and discussions. This work is supported by Grants-in-Aid for Scientific Research (\#23103005, 15H02065) from MEXT of Japan.

\appendix
\section{Terminal Velocity}
For particles much larger than the mean free path of the molecules in air, 
the terminal velocity of a particle under gravity $g$ is given by 
\begin{equation}
v_{\rm t}(r) =\sqrt{ \frac{8gr\rho_{\rm p}}{3C_{\rm D}\rho_{\rm a}} },
\label{eq:vt}
\end{equation}
where $\rho_{\rm a}$ is the mass density of the atmosphere, and $C_{\rm D}$ is the drag coefficient
determined by the local Reynolds number of the particle,
\begin{equation}
N_{\rm Re} \equiv \frac{2rv_{\rm t}(r)\rho_{\rm a}}{\eta},
\label{eq:NRe}
\end{equation}
where $\eta$ is the dynamic viscosity of the atmosphere.

\begin{figure}
\includegraphics[clip,width=\hsize]{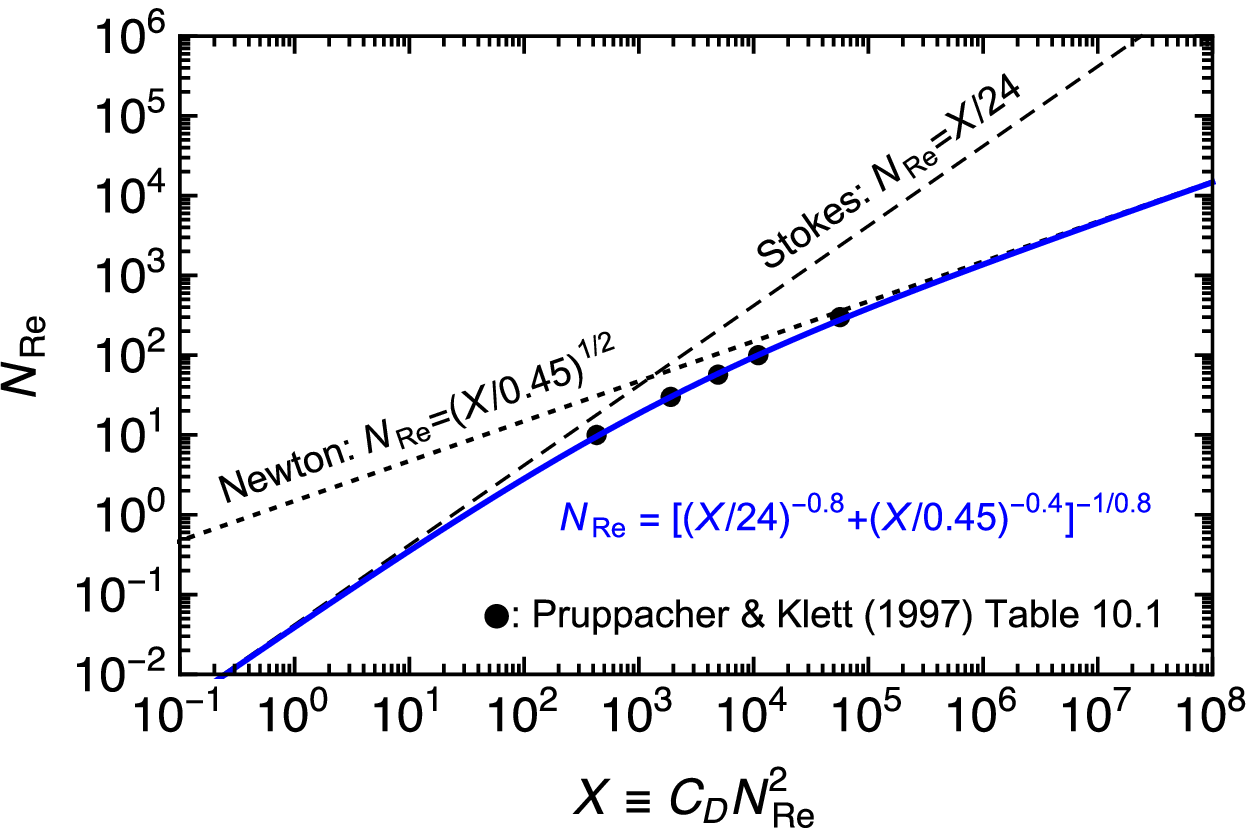}
\caption{Particle Reynolds number $N_{\rm Re}$ versus $X\equiv C_{\rm D}N_{\rm Re}^2$. 
The dashed and dotted lines indicate the Stokes and Newton drag laws, respectively, 
and the solid circles show the data for rigid spheres from Table 10.1 of \citet{Pru97}. 
The solid line shows our fit, Equation~\eqref{eq:NRe_fit}.
}
\label{fig:NRe}
\end{figure}
Because $N_{\rm Re}$ depends on $v_{\rm t}$, 
one needs to solve Equations~\eqref{eq:vt} and \eqref{eq:NRe}, 
together with the relation between $C_{\rm D}$ and $N_{\rm Re}$,
to obtain $v_{\rm t}$ as a function of $r$.
Following \citet{AM01}, we do so by introducing the quantity 
\begin{equation}
X \equiv C_{\rm D}N_{\rm Re}^2 = \frac{32gr^3 \rho_{\rm a}\rho_{\rm p}}{3\eta^2},
\label{eq:X}
\end{equation}
which does not involve $v_{\rm t}$, 
and analytically express $N_{\rm Re}$ as a function of $X$. 
We require the function to reproduce Stokes' and Newton's drag laws, 
$C_{\rm D} = 24/N_{\rm Re}$ ($N_{\rm Re} = X/24$) 
and $C_{\rm D} = 0.45$ ($N_{\rm Re} = \sqrt{X/0.45}$),
in the limits of $N_{\rm Re} \ll 1$ and $N_{\rm Re} \gg 1000$, respectively. 
To determine the functional form in the intermediate regime $1 \la N_{\rm Re} \la 1000$,
we use the data for rigid spheres from Table 10.1 of \citet{Pru97}. 
We find that the function
\begin{equation}
N_{\rm Re} = \left[\left(\frac{X}{24}\right)^{-0.8} + \left(\frac{X}{0.45}\right)^{-0.4} \right]^{-1/0.8}
\label{eq:NRe_fit}
\end{equation}
well reproduces the data points in the intermediate regime and the correct limiting behaviors 
at $N_{\rm Re} \ll 1$ and $N_{\rm Re} \gg 1000$ (see Figure~\ref{fig:NRe}).
Substituting this expression and Equation~\eqref{eq:X} into Equation~\eqref{eq:NRe}, 
we obtain the analytic expression for $v_{\rm t}$ as a function of $r$,
\begin{equation}
v_{\rm t} = \frac{2gr^2 \rho_{\rm p}}{9\eta}
\left[ 1+\left(\frac{0.45gr^3\rho_{\rm a}\rho_{\rm p}}{54\eta^2}\right)^{0.4} \right]^{-1.25}.
\label{eq:vt_fit}
\end{equation}

If the particles are small the atmospheric mean free path, one should use 
the Epstein law
\begin{equation}
v_{\rm t}(r)=\frac{g\rho_{\rm p}}{3C_{\rm s}\rho_{\rm a}}r
\end{equation}
instead of Equation~\eqref{eq:vt_fit}. 
This is the case for high-altitude clouds in exoplanets (see Section~\ref{sec:cond}).



\begin{thebibliography}{}
\bibitem[Ackerman \& Marley(2001)]{AM01} Ackerman, A. S., \& Marley, M. S. 2001, \apj, 556, 872
\bibitem[Ackerman \& Marley(2013)]{AM13} Ackerman, A.~S., \& Marley, M.~S.\ 2013, \apj, 765, 75 
\bibitem[Atreya et al.(2005)]{Atreya05} Atreya, S. K., \& Wong, A.-S. 2005, \ssr, 116, 121
\bibitem[Bean et al.(2010)]{Bean10} Bean, J.~L., Miller-Ricci Kempton, E., \& Homeier, D.\ 2010, \nat, 468, 669
\bibitem[Birnstiel et al.(2012)]{Birnstiel+12} Birnstiel, T., Klahr, H., \& Ercolano, B.\ 2012, \aap, 539, A148 
\bibitem[Blum \& Wurm(2000)]{Blum00} Blum, J., \& Wurm, G.\ 2000, \icarus, 143, 138 
\bibitem[Buenzli et al.(2012)]{Buenzli+12} Buenzli, E., Apai, D., Morley, C.~V., et al.\ 2012, \apjl, 760, L31 
\bibitem[Burgasser et al.(2002)]{Burgasser+02} Burgasser, A.~J., Marley, M.~S., Ackerman, A.~S., et al.\ 2002, \apjl, 571, L151 
\bibitem[Carlson et al.(1994)]{Carlson94} Carlson, B.~E., Lacis, A.~A., \& Rossow, W.~B.\ 1994, \jgr, 99, 14 
\bibitem[Cooper et al.(2003)]{Cooper03} Cooper, C.~S., Sudarsky, D., Milsom, J.~A., Lunine, J.~I., \& Burrows, A.\ 2003, \apj, 586, 1320 
\bibitem[Crossfield et al.(2013)]{Crossfield+13} Crossfield, I.~J.~M., Barman, T., Hansen, B.~M.~S., \& Howard, A.~W.\ 2013, \aap, 559, A33 
\bibitem[Dragomir et al.(2015)]{Dragomir15} Dragomir, D., Benneke, B., Pearson, K.~A., et al.\ 2015, \apj, 814, 102 
\bibitem[Ehrenreich et al.(2014)]{Enrenreich14} Ehrenreich, D., Bonfils, X., Lovis, C., et al.\ 2014, \aap, 570, A89 
\bibitem[Ferrier(1994)]{Ferrier94} Ferrier, B. S. 1994, J. Atmos. Sci., 51, 249
\bibitem[Fortney(2005)]{Fortney05} Fortney, J.~J.\ 2005, \mnras, 364, 649 
\bibitem[Guillot et al.(2014)]{Guillot14} Guillot, T., \& Ormel, C. W. 2014, \aap, 572, A72
\bibitem[Hansen(1979)]{Hansen79} Hansen, C. F. 1979, NASA Technical Memo 78556
\bibitem[Helling \& Woitke(2006)]{Helling06} Helling, Ch., \& Woitke, P. 2006, \aap, 455, 325 
\bibitem[Helling et al.(2008)]{Helling08} Helling, C., Woitke, P., \& Thi, W.-F. 2008, \aap, 485, 547
\bibitem[Helling et al.(2016)]{Helling16} Helling, C., Lee, G., Dobbs-Dixon, I., et al.\ 2016, \mnras, 460, 855 
\bibitem[Homann et al.(2016)]{Homann+16} Homann, H., Guillot, T., Bec, J., et al.\ 2016, \aap, 589, A129
\bibitem[Kataoka et al.(2013)]{Kataoka13} Kataoka, A., Tanaka, H., Okuzumi, S., \& Wada, K.\ 2013, \aap, 554, A4 
\bibitem[Kokhanovsky(2004)]{Kokhanovsky04} Kokhanovsky, A. 2004, Earth-Science Reviews, 64, 189
\bibitem[Knutson et al.(2014a)]{Knutson14a} Knutson, H. A., Benneke, B., Deming, D., \& Homeier, D. 2014, Nature, 505, 66
\bibitem[Knutson et al.(2014b)]{Knutson14b} Knutson, H.~A., Dragomir, D., Kreidberg, L., et al.\ 2014, \apj, 794, 155 
\bibitem[Kreidberg et al.(2014)]{Kreidberg14} Kreidberg, L., Bean, J. L., D\'{e}sert, J.-M., et al. 2014, Nature, 505, 69
\bibitem[Krijt et al.(2016)]{Krijt16} Krijt, S., Ormel, C.~W., Dominik, C., \& Tielens, A.~G.~G.~M.\ 2016, \aap, 586, A20 
\bibitem[Lee et al.(2015)]{Lee15} Lee, G., Helling, C., Dobbs-Dixon, I., \& Juncher, D.\ 2015, \aap, 580, A12 
\bibitem[Marley et al.(2002)]{Marley+02} Marley, M.~S., Seager, S., Saumon, D., et al.\ 2002, \apj, 568, 335 
\bibitem[Marley et al.(2013)]{Marley13} Marley, M.~S., Ackerman, A.~S., Cuzzi, J.~N., \& Kitzmann, D.\ 2013, Comparative Climatology of Terrestrial Planets, 367 
\bibitem[Morley et al.(2013)]{Morley13} Morley, C. V., Fortney, J. J., Kempton, E. M.-R., et al. 2013, \apj, 775, 33
\bibitem[Morley et al.(2015)]{Morley15} Morley, C. V., Fortney, J. J., Marley, M. S., et al. 2015, \apj, 815, 110
\bibitem[Okuzumi et al.(2009)]{Okuzumi09} Okuzumi, S., Tanaka, H., \& Sakagami, M.-a.\ 2009, \apj, 707, 1247 
\bibitem[Okuzumi et al.(2012)]{Okuzumi+12} Okuzumi, S., Tanaka, H., Kobayashi, H., \& Wada, K.\ 2012, \apj, 752, 106
\bibitem[Ormel et al.(2014)]{Ormel14} Ormel, C. W. 2014, \apj, 789, L18
\bibitem[Ormel et al.(2007)]{Ormel07} Ormel, C.~W., Spaans, M., \& Tielens, A.~G.~G.~M.\ 2007, \aap, 461, 215
\bibitem[Pont et al.(2013)]{Pont13} Pont, F., Sing, D.~K., Gibson, N.~P., et al.\ 2013, \mnras, 432, 2917 
\bibitem[Pruppacher \& Klett(1997)]{Pru97} Pruppacher, H. R., \& Klett, J. D. 1997, Microphysics of Clouds and Precipitation (Dordrecht: Kluwer)
\bibitem[Ragent et al.(1998)]{Ragent98} Ragent, B., Colburn, D. S., Rages, K. A., et al. 1998, \jgr, 103, 22891
\bibitem[Rauber et al.(2007)]{Rauber07} Rauber, R.~M., Stevens, B., Ochs, H.~T., et al. 2007, Bull. Amer. Meteor. Soc., 88, 1912
\bibitem[Rogers \& Yau(1989)]{Rogers89} Rogers, R., \& Yau, M. 1989, A Short Course in Cloud Physics (3rd ed.; Oxford: Butterworth-Heinemann)
\bibitem[Rossow(1978)]{Rossow78} Rossow, W. B. 1978, Icarus, 36, 1
\bibitem[Sato et al.(2016)]{Sato+16} Sato, T., Okuzumi, S., \& Ida, S. 2016, \aap, 589, A15
\bibitem[Saumon \& Marley(2008)]{Saumon&Marley08} Saumon, D., \& Marley, M.~S.\ 2008, \apj, 689, 1327-1344 
\bibitem[Seager \& Sasselov(2000)]{Seager00} Seager, S., \& Sasselov, D.~D.\ 2000, \apj, 537, 916 
\bibitem[Seinfeld \& Pandis(2006)]{Seinfeld06} Seinfeld, J. H., \& Pandis, S. N. 2006, Atmospheric Chemistry and Physics:
From Air Pollution to Climate Change (2nd ed.; New Jersey: Wiley)
\bibitem[Seiff et al.(1998)]{Seiff98} Seiff, A., Kirk, D. B., Knight, T. C. D., et al. 1998, \jgr, 103, 889, 22822
\bibitem[Sing et al.(2015)]{Sing15} Sing, D.~K., Wakeford, H.~R., Showman, A.~P., et al.\ 2015, \mnras, 446, 2428 
\bibitem[Sing et al.(2016)]{Sing16} Sing, D.~K., Fortney, J.~J., Nikolov, N., et al.\ 2016, \nat, 529, 59 
\bibitem[Slinn(1974)]{Slinn74} Slinn, W. G. N. 1974, Proceedings of the USAEC Symposium
\bibitem[Stevenson et al.(2016)]{Stevenson+16} Stevenson, K.~B., Bean, J.~L., Seifahrt, A., et al.\ 2016, \apj, 817, 141 
\bibitem[Sugiyama et al.(2014)]{Sugiyama14} Sugiyama, K., Nakajima, K., Odaka, M., Kuramoto, K., \& Hayashi, Y.-Y.\ 2014, \icarus, 229, 71 
\bibitem[vanZanten et al.(2011)]{van11} VanZanten, M. C., Stevens, B., Nuijens, L., et al. 2011, J. Advances in Modelling Earth System, 3, M06001
\bibitem[Weidenscilling \& Lewis(1973)]{WL73} Weidenschilling, S. J., \& Lewis, J. S. 1973, Icarus, 20, 465
\bibitem[West et al.(1986)]{West86}West, R. A., Strobel, D. F., \& Tomasko, M. G. 1986, Icarus, 65, 161
\bibitem[Woitke \& Helling(2003)]{Woitke03} Woitke, P., \& Helling, C. 2003, \aap, 399, 297
\bibitem[Woitke \& Helling(2004)]{Woitke04} Woitke, P., \& Helling, C. 2004, \aap, 414, 335
\bibitem[Wong et al.(2004)]{Wong04} Wong, M. H., Mahaffy, P. R., Atreya, S. K., Niemann, H. B., \& Owen, T. C. 2004, Icarus, 171, 153
\bibitem[Yang et al.(2015)]{Yang+15} Yang, H., Apai, D., Marley, M.~S., et al.\ 2015, \apjl, 798, L13 
\bibitem[Yang et al.(2016)]{Yang+16} Yang, H., Apai, D., Marley, M.~S., et al.\ 2016, \apj, 826, 8
\bibitem[Ziegler(1985)]{Ziegler85} Ziegler, C. L., 1985, J. Atmos. Sci., 42, 1487
\bibitem[Zsom et al.(2012)]{Zsom12} Zsom, A., Kaltenegger, L., \& Goldblatt, C. 2012, Icarus, 221, 603.

\end{thebibliography}
\end{document}